\documentclass[12pt]{iopart}

\usepackage[utf8]{inputenc}
\usepackage[T1]{fontenc}
\usepackage[english]{babel}

\usepackage[pdftex]{graphicx}
\usepackage{bbm,amsthm,amsfonts,amssymb,amsopn,amsmath}
\usepackage{dsfont}
\usepackage{mathtools}
\usepackage{url,hyperref,cite}
\usepackage{pgf,tikz}
\usepackage{iopams}
\usepackage{calc,xstring,ifthen}
\usepackage{pdftexcmds,adjustbox}
\usetikzlibrary{arrows,calc,matrix,backgrounds,fit,positioning,decorations.pathmorphing,decorations.markings}
\usepackage{bbm,bm}
\usepackage{cleveref}

\makeatletter
\renewcommand*\env@matrix[1][\arraystretch]{%
  \edef\arraystretch{#1}%
  \hskip -\arraycolsep
  \let\@ifnextchar\new@ifnextchar
  \array{*\c@MaxMatrixCols c}}
\makeatother
\renewcommand{\tr}{\mathrm{tr}}

\def\be{\begin{equation}}
\def\ee{\end{equation}}
\def\bra#1{\mathinner{\langle{#1}|}}
\def\ket#1{\mathinner{|{#1}\rangle}}

\newcommand{\bea}{\begin{eqnarray}}
\newcommand{\eea}{\end{eqnarray}}
\newcommand{\bbra}[1]{({#1}|}
\newcommand{\kket}[1]{|{#1})}
\newcommand{\bbrakket}[2]{(#1 \vert #2 )}

\newcommand{\braket}[2]{\langle #1 \vert #2 \rangle}

\newcommand{\ave}[1]{{\langle #1\rangle}}

\newcommand{\ii}{ {\rm i} }

\newcommand{\RR}{\mathbb{R}}

\renewcommand{\vec}[1]{{\mathbf{#1}}}
\newcommand{\unit}{\mathbbm{1}}
\definecolor{full}{rgb}{0,0,0}
\definecolor{old}{rgb}{1,1,1}
\definecolor{halfborder}{rgb}{0.5,0.5,0.5}
\definecolor{half}{rgb}{0.9,0.9,0.9}
\definecolor{border}{rgb}{0.3,0.3,0.3}
\def\a{0.2}
\def\ba{0.45}

\newcommand\emptyrectangle[2]{
  \draw[border] ({\a*(#1)},{\a*(#2-1)})  -- ({\a*(#1+1)},{\a*(#2)})  -- ({\a*(#1)},{\a*(#2+1)})  
  -- ({\a*(#1-1)},{\a*(#2)})  -- cycle;
}
\newcommand\fullrectangle[2]{
  \draw[border,fill=full] ({\a*(#1)},{\a*(#2-1)})  -- ({\a*(#1+1)},{\a*(#2)})  -- ({\a*(#1)},{\a*(#2+1)})  
  -- ({\a*(#1-1)},{\a*(#2)})  -- cycle;
}
\newcommand\rectangle[3]{
  \ifthenelse{\equal{#3}{1}}{\fullrectangle{#1}{#2}}{\emptyrectangle{#1}{#2}};
}
\newcommand\bigdashedrectangle[2]{
  \draw[white,thick] ({\ba*(#1)},{\ba*(#2-1)})  -- ({\ba*(#1+1)},{\ba*(#2)})  -- ({\ba*(#1)},{\ba*(#2+1)})  
  -- ({\ba*(#1-1)},{\ba*(#2)})  -- cycle;
  \draw[border,thin,dashed] ({\ba*(#1)},{\ba*(#2-1)})  -- ({\ba*(#1+1)},{\ba*(#2)})  -- ({\ba*(#1)},{\ba*(#2+1)})  
  -- ({\ba*(#1-1)},{\ba*(#2)})  -- cycle;
}
\newcommand\bigtextrectangle[4]{
  \draw[border,fill=#4] ({\ba*(#1)},{\ba*(#2-1)})  -- ({\ba*(#1+1)},{\ba*(#2)})  -- ({\ba*(#1)},{\ba*(#2+1)})  
  -- ({\ba*(#1-1)},{\ba*(#2)})  -- cycle;
  \node at ({\ba*(#1)},{\ba*(#2)}) {\scalebox{0.7}{#3}}
}
\newcommand\textrectangle[3]{
  \draw[border,fill=half] ({\a*(#1)},{\a*(#2-1)})  -- ({\a*(#1+1)},{\a*(#2)})  -- ({\a*(#1)},{\a*(#2+1)})  
  -- ({\a*(#1-1)},{\a*(#2)})  -- cycle;
  \node at ({\a*(#1)},{\a*(#2)}) {\scalebox{0.7}{#3}}
}

\newcommand\configThreeLeft[3]{
  \rectangle{(0.5}{(-1)}{#1};
  \rectangle{(-0.5}{(0)}{#2};
  \rectangle{(0.5}{(1)}{#3};
}
\newcommand\configThreeRight[3]{
  \rectangle{(-0.5}{(-1)}{#1};
  \rectangle{(0.5}{(0)}{#2};
  \rectangle{(-0.5}{(1)}{#3};
}
\newcommand\configFourLeft[4]{
  \rectangle{(0.5}{(-1)}{#1};
  \rectangle{(-0.5}{(0)}{#2};
  \rectangle{(0.5}{(1)}{#3};
  \rectangle{(-0.5}{(2)}{#4};
}
\newcommand\configFourRight[4]{
  \rectangle{(-0.5}{(-1)}{#1};
  \rectangle{(0.5}{(0)}{#2};
  \rectangle{(-0.5}{(1)}{#3};
  \rectangle{(0.5}{(2)}{#4};
}
\newcommand\rightSaw[4]{
  \begin{tikzpicture}[baseline={([yshift=-0.5ex]current bounding box.center)}, scale=0.6] 
    \rectangle{0}{-0.5}{#1};
    \rectangle{1}{0.5}{#2};
    \rectangle{2}{-0.5}{#3};
    \rectangle{3}{0.5}{#4};
  \end{tikzpicture}%
}
\newcommand\leftSaw[4]{
  \begin{tikzpicture}[baseline={([yshift=-0.5ex]current bounding box.center)}, scale=0.6] 
    \rectangle{0}{0.5}{#1};
    \rectangle{1}{-0.5}{#2};
    \rectangle{2}{0.5}{#3};
    \rectangle{3}{-0.5}{#4};
  \end{tikzpicture}%
}
\newcommand\rightsSaw[2]{
  \begin{tikzpicture}[baseline={([yshift=-0.5ex]current bounding box.center)}, scale=0.6] 
    \rectangle{0}{-0.5}{#1};
    \rectangle{1}{0.5}{#2};
  \end{tikzpicture}%
}
\newcommand\leftsSaw[2]{
  \begin{tikzpicture}[baseline={([yshift=-0.5ex]current bounding box.center)}, scale=0.6] 
    \rectangle{0}{0.5}{#1};
    \rectangle{1}{-0.5}{#2};
  \end{tikzpicture}%
}
\newcommand\rcaTextRule[4]{
  \begin{tikzpicture}[baseline={([yshift=-0.1ex]current bounding box.center)},scale=1.8]
    \rectangle{0}{0}{#1};
    \rectangle{1}{-1}{#2};
    \rectangle{2}{0}{#3};
    \rectangle{1}{1}{#4};
    \node[text=halfborder] at ({\a*(0)},{\a*(0)}) {\scalebox{0.7}{$s_1$}};
    \node[text=halfborder] at ({\a*(1)},{\a*(-1)}) {\scalebox{0.7}{$s_2$}};
    \node[text=halfborder] at ({\a*(2)},{\a*(0)}) {\scalebox{0.7}{$s_3$}};
    \node[text=halfborder] at ({\a*(1)},{\a*(1)}) {\scalebox{0.7}{$s_2^{\prime}$}};
  \end{tikzpicture}
}
\newcommand\rcaRule[4]{
  \begin{tikzpicture}[baseline={([yshift=-0.1ex]current bounding box.center)},scale=1.8]
    \rectangle{0}{0}{#1};
    \rectangle{1}{-1}{#2};
    \rectangle{2}{0}{#3};
    \rectangle{1}{1}{#4};
  \end{tikzpicture}
}
\begin{document}

\title{Matrix product state of multi-time correlations}
\author{Katja~Klobas$^1$, Matthieu~Vanicat$^1$, Juan~P.~Garrahan$^{2,3}$ and Toma\v z~Prosen$^1$}
\address{$^1$ Department of Physics, Faculty of Mathematics and Physics, University of Ljubljana, Ljubljana, Slovenia\\
         $^2$ School of Physics and Astronomy, University of Nottingham, Nottingham NG7~2RD, United Kingdom\\
         $^3$ Centre for the Mathematics and Theoretical Physics of Quantum Non-equilibrium Systems,University of Nottingham,
         Nottingham NG7~2RD, United Kingdom}

\begin{abstract}
 For an interacting spatio-temporal lattice system we introduce a formal way of
 expressing multi-time correlation functions of local observables located at
 the same spatial point with a~\emph{time state}, i.e.\ a~statistical distribution of
 configurations observed along a time lattice. Such a time state is defined
 with respect to a particular equilibrium state that is invariant under space
 and time translations. The concept is developed within the \emph{Rule 54}
 reversible cellular automaton, for which we explicitly construct a~matrix
 product form of the time state, with matrices that act on the $3$-dimensional
 auxiliary space. We use the matrix-product state to express equal-space
 time-dependent density-density correlation function, which, for special
 maximum-entropy values of equilibrium parameters, agrees with the previous
 results.  Additionally, we obtain an explicit expression for the
 probabilities of observing all multi-time configurations, which enables us to
 study distributions of times between consecutive excitations and prove the
 absence of decoupling of timescales in the Rule 54 model.
\end{abstract}

\section{Introduction}

When discussing locally interacting models in statistical mechanics,
the central problem is the computation of time-dependent correlation functions of local
observables, especially when studying non-equilibrium dynamics, transport,
thermalization and alike. Multi-time correlation functions, in particular, are
crucial for the understanding of complex dynamics that
occurs in systems displaying metastability, such as glass formers (for reviews,
see e.g.\ \cite{Chandler2010,Binder2011,Biroli2013}). However, obtaining an
exact form or at least asymptotic behaviour of such correlation functions is
usually out of the scope of the current analytical methods, even for the
simplest interacting systems. Sometimes even partial knowledge of multi-time
correlation functions of local observables located at the same point can
provide useful insights. For example, when studying the transport of conserved
quantities, knowing the multi-time equal-space density-density correlation
function provides the scaling behaviour of the central (heat) peak within the
hydrodynamic description~\cite{spohn2014nonlinear}. On other hand, multi-time
equal-space correlation functions can be viewed as expectation values of a
specific class of observables that are nonlocal in time, but local in space.

\begin{figure}
  \centering
  \begin{tikzpicture}
    % Example of dynamics:
    \draw [gray,densely dashed] (-1.25,0.25) -- (-2.5,1.5);
    \draw [gray,densely dashed] (-3.75,0.25) -- (-2.5,1.5);
    \draw [gray,densely dashed] (-2.5,1.5) -- (-2.5,1.75);
    \draw [gray,densely dashed] (-2.5,1.75) -- (-4,3.25);
    \draw [gray,densely dashed] (0.75,0.25) -- (0.75,0.5);
    \draw [gray,densely dashed] (0.75,0.5) -- (-0.125,1.375) -- (-1.25,0.25);
    \draw [gray,densely dashed] (-0.125,1.625) -- (-1.375,2.875) -- (-2.5,1.75);
    \draw [gray,densely dashed] (-1.375,2.875) -- (-1.375,3.125);
    \draw [gray,densely dashed] (1.25,0.25) -- (0.875,0.625) -- (0.75,0.5);
    \draw [gray,densely dashed] (0.875,0.625) -- (0.875,0.875);
    \draw [gray,densely dashed] (-0.125,1.375) -- (-0.125,1.625) -- (0,1.75);
    \draw [gray,densely dashed] (0,1.75) -- (0.875,0.875);
    \draw [gray,densely dashed] (0,1.75) -- (0,2);
    \draw [gray,densely dashed] (-2.75,4.5) -- (-1.375,3.125) -- (-1.25,3.25) -- (0,2);
    \draw [gray,densely dashed] (-1.25,3.25) -- (-1.25,3.5) -- (-2.25,4.5);
    \draw [gray,densely dashed] (-1.25,3.5) -- (-0.25,4.5);
    \draw [gray,densely dashed] (2.5,0.25) -- (3.25,1) -- (4,0.25);
    \draw [gray,densely dashed] (3.25,1) -- (3.25,1.25) -- (4,2);
    \draw [gray,densely dashed] (3.25,1.25) -- (2.25,2.25) -- (0.875,0.875); 
    \draw [gray,densely dashed] (2.25,2.25) -- (2.25,2.5) -- (1.375,3.375) -- (0,2);
    \draw [gray,densely dashed] (1.375,3.375) -- (1.375,3.625) -- (2.25,4.5);
    \draw [gray,densely dashed] (1.375,3.625) -- (0.5,4.5);
    \draw [gray,densely dashed] (2.25,2.5) -- (4,4.25);
    % x and t axes:
    \draw [->,>=stealth,densely dotted,thick] (0,0) -- (0,5);
    \draw [->,>=stealth,densely dotted,thick] (-5,0) -- (5,0);
    \node at (4.75,0.25) {$x$};
    \node at (0.25,4.75) {$t$};
    % equilibrium state
    \coordinate (A) at (-4,0.25);
    \coordinate (B) at (4,0.25);
    \coordinate (C) at (4,-0.25);
    \coordinate (D) at (-4,-0.25);
    \filldraw[thick,
      draw=black,fill=halfborder,rounded corners=0.2,
      decoration={snake,segment length=1.081mm,amplitude=0.6mm}
    ]
    (A) -- (B)
    decorate {-- (C)}  -- (D)
    decorate {-- (A)};
    \node at (0,0) {equilibrium state $\vec{p}$};
    % Vertical state Vertical state
    \filldraw [thick,rounded corners=1,draw=black,fill=half] (0.1,0.5) -- (0.1,4.5) -- (-0.1,4.5) -- (-0.1,0.5) -- cycle;
    \filldraw (0,0.7) circle [radius=0.045];
    \filldraw (0,1.3) circle [radius=0.045];
    \filldraw (0,1.7) circle [radius=0.045];
    \filldraw (0,2.7) circle [radius=0.045];
    \filldraw (0,3.1) circle [radius=0.045];
    \filldraw (0,3.6) circle [radius=0.045];
    \filldraw (0,4.3) circle [radius=0.045];
    \node at (-0.5,0.7) {$b_0$};
    \node at (-0.5,1.3) {$b_1$};
    \node at (-0.5,3.2) {$\vdots$};
    \node at (-0.5,4.3) {$b_k$};
    \node at (0.375,1.875) {$\vec{q}$};
  \end{tikzpicture}
  \caption{\label{fig:schPicture}
    Schematic picture of the setup. We are
    interested in probability distribution~$\vec{q}$ of the
    configurations~$(b_0,b_1,\cdots,b_k)$ at the center of the chain, $x=0$, and
    different times $t$, while the system is in the equilibrium state~$\vec{p}$.
    Intuitively, we can understand this as keeping track of the particles that
    pass through the origin. Since they are interacting with each other, this is
    generally not a~trivial task and the computational complexity of~$\vec{q}$ might still
    be growing exponentially with time. 
    }
\end{figure}
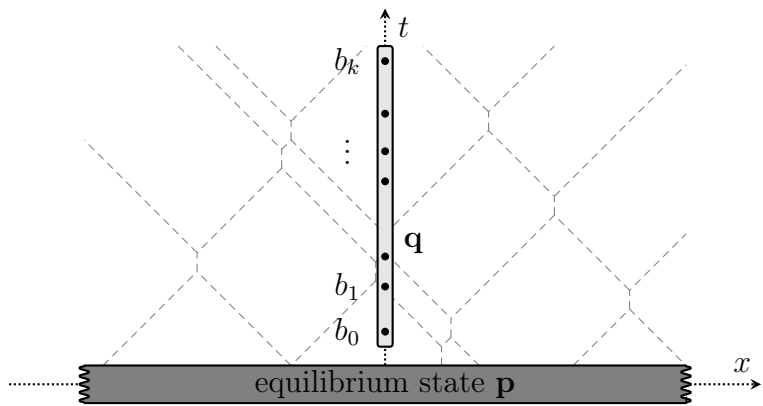

Specifically, we will be interested in the multi-point correlation functions of
single-site (ultralocal) observables $a_j$ located at the same spatial point
$x=0$ but at different times $t_j$,
\begin{eqnarray}\label{eq:expVal} 
  C_{a_1,a_2,\ldots,a_k}(t_1,t_2,\ldots,t_k)=
  \ave{a_1(0,t_1)a_2(0,t_2)\cdots a_k(0,t_k)}_{\vec{p}},
\end{eqnarray}
where $\vec{p}$ denotes an equilibrium state (which is, by definition,
time-translation invariant).  The computational complexity of the time evolution
of local observables in interacting systems generically grows exponentially even
within classical mechanics~\footnote{By computational complexity we refer to the
cost of computing the propagated observable exactly, not
approximating it (numerically).}, but since all of them are positioned at the
same point, one expects that a lot of information contained in the time
propagated observables~$a_j(x,t)$ is not necessary for knowing the expectation
value (\ref{eq:expVal}). Therefore we propose that, at least in certain cases,
the computational complexity could be reduced by introducing a~\emph{time state
$\vec{q}(\vec{p})$}, i.e.\ the~probability distribution of observing
\emph{configurations in time} $(b_0, b_1, b_2,\ldots, b_k)$ if the system is in
the equilibrium state~$\vec{p}$, as is schematically shown in
Figure~\ref{fig:schPicture}. Here $b_j$ label a complete set of ultralocal
observables.

In the paper we show that it is possible to construct an explicit nontrivial
matrix-product representation of such states in the case of Rule 54 reversible
cellular automaton (RCA54), introduced in~\cite{bobenko1993two}. RCA54 is a
classical deterministic model in discrete spacetime of particles that move with
constant velocities and interact pairwise by shifting for one site when
scattering.  The model is also closely related to the automaton ERCA 250R, as
introduced by Takesue~\cite{Takesue}, and to the Floquet version of
Fredrickson-Andersen
model~\cite{gopalakrishnan1,gopalakrishnan2,gopalakrishnan3}.  The simplicity
of the model allows for a lot of exact results, such as the
non-equilibrium steady state of the system coupled to stochastic
reservoirs~\cite{prosenMejiaMonasterioCA54,inoueTakesueCA54}, as well the
computation of
the most relevant (long lived) decay modes~\cite{prosenBucaCA54}, and exact
treatment of large deviation statistics~\cite{bucaetalLargeDev}.
An explicit matrix-product representation of the complete  time evolution of
local observables was found in~\cite{TMPA2018}, and was afterwards generalized
to the time evolution of local density matrices in the quantum version of the
model~\cite{alba2019RCA54}.

The paper is structured as follows; in section~\ref{sec:rca54} we define the
model and introduce a~two-parameter family of equilibrium states. They
are expressed in terms of a matrix-product ansatz (MPA), with matrices that obey
the cubic cancellation relation introduced in~\cite{prosenBucaCA54}. We express
asymptotic probabilities of observing finite subconfigurations in the limit of
large system sizes and show that the probabilities of observing solitons are
statistically independent. Section~\ref{sec:vState} contains the main result of
the paper, namely the explicit construction of the~time state. Due to the
statistical independence of solitons, the construction reduces to the problem
of keeping track of all the quasi-particles that pass
through the site $0$. This is achieved by introducing matrices that act on
a~$3$-dimensional auxiliary space and expressing the time state in an~MPA form.
In sections~\ref{sec:2pointCorrs} and~\ref{sec:timeConfProbs} we briefly
discuss examples of correlation functions and expectation values that can be
exactly expressed using the MPA form of the time state. Finally, in
section~\ref{sec:conclusion} we finish with closing remarks. 

\section{Rule 54 reversible cellular automaton}\label{sec:rca54}
\subsection{Definition of the dynamics}
\begin{figure}
  \begin{equation*}
    \begin{tikzpicture}[baseline={([yshift=0.5ex]current bounding box.center)}]
    \bigtextrectangle{0}{0}{$s_{x}^{t}$}{half};
    \bigtextrectangle{1}{-1}{$s_{x+1}^{t}$}{half};
    \bigtextrectangle{2}{0}{$s_{x+2}^{t}$}{half};
    \bigtextrectangle{3}{-1}{$s_{x+3}^{t}$}{half};
  \end{tikzpicture}\longrightarrow\ 
  \begin{tikzpicture}[baseline={([yshift=-0.75ex]current bounding box.center)}]
    \bigdashedrectangle{1}{-1};
    \bigdashedrectangle{3}{-1};
    \bigtextrectangle{0}{0}{$s_{x}^{t+1}$}{half};
    \bigtextrectangle{1}{1}{$s_{x+1}^{t+1}$}{halfborder};
    \bigtextrectangle{2}{0}{$s_{x+2}^{t+1}$}{half};
    \bigtextrectangle{3}{1}{$s_{x+3}^{t+1}$}{halfborder};
  \end{tikzpicture}
\longrightarrow\ 
\begin{tikzpicture}[baseline={([yshift=-2ex]current bounding box.center)}]
    \bigdashedrectangle{0}{0};
    \bigdashedrectangle{1}{-1};
    \bigdashedrectangle{2}{0};
    \bigdashedrectangle{3}{-1};
    \bigtextrectangle{0}{2}{$s_{x}^{t+2}$}{halfborder};
    \bigtextrectangle{1}{1}{$s_{x+1}^{t+2}$}{halfborder};
    \bigtextrectangle{2}{2}{$s_{x+2}^{t+2}$}{halfborder};
    \bigtextrectangle{3}{1}{$s_{x+3}^{t+1}$}{halfborder};
  \end{tikzpicture}
  \end{equation*}
  \caption{\label{fig:lattGeo}
    Schematic representation of the lattice geometry
    and time evolution. Only half of the sites are
    being updated at the same time~\eqref{eq:timeProp}, therefore it is convenient to imagine
    the chain to have a zig-zag shape. At each time step,
    the sites positioned at the bottom (e.\ g.\ $s_{x+1}^{t}$ and $s_{x+3}^t$ on the left
    most picture) are propagated, while the top sites ($s_x^t$ and $s_{x+2}^t$)
    do not change. The new value depends on the values of the three consecutive sites
    as described in~\eqref{eq:timeProp} and~\eqref{eq:rcarules}.
  }
\end{figure}
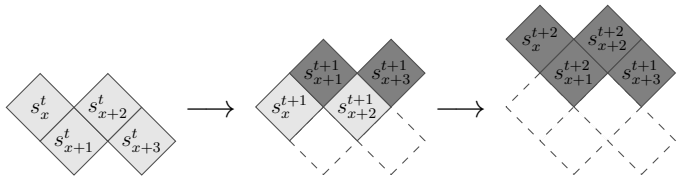

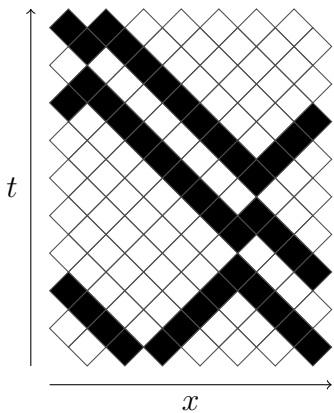
\begin{figure}
  \centering
  \begin{tikzpicture}[scale=1.25]
    \draw[->] ({-2*(\a)},{-2*(\a)}) -- ({-2*(\a)},{17*(\a)});
    \node at ({-3*\a},{7.5*(\a)}) {$t$};
    \node at ({6.5*\a},{-4*(\a)}) {$x$};
    \draw[->] ({-1*(\a)},{-3*(\a)}) -- ({14*(\a)},{-3*(\a)});
    \rectangle{0}{0}{0};
    \rectangle{1}{-1}{0};
    \rectangle{2}{0}{1};
    \rectangle{3}{-1}{1};
    \rectangle{4}{0}{0};
    \rectangle{5}{-1}{1};
    \rectangle{6}{0}{1};
    \rectangle{7}{-1}{0};
    \rectangle{8}{0}{0};
    \rectangle{9}{-1}{0};
    \rectangle{10}{0}{0};
    \rectangle{11}{-1}{0};
    \rectangle{12}{0}{1}
    \rectangle{13}{-1}{1}
    \rectangle{0}{2}{1};
    \rectangle{1}{1}{1};
    \rectangle{2}{2}{0};
    \rectangle{3}{1}{0};
    \rectangle{4}{2}{0};
    \rectangle{5}{1}{0};
    \rectangle{6}{2}{0};
    \rectangle{7}{1}{1};
    \rectangle{8}{2}{1};
    \rectangle{9}{1}{0};
    \rectangle{10}{2}{1};
    \rectangle{11}{1}{1};
    \rectangle{12}{2}{0};
    \rectangle{13}{1}{0};
    \rectangle{0}{4}{0};
    \rectangle{1}{3}{0};
    \rectangle{2}{4}{0};
    \rectangle{3}{3}{0};
    \rectangle{4}{4}{0};
    \rectangle{5}{3}{0};
    \rectangle{6}{4}{0};
    \rectangle{7}{3}{0};
    \rectangle{8}{4}{0};
    \rectangle{9}{3}{1};
    \rectangle{10}{4}{0};
    \rectangle{11}{3}{0};
    \rectangle{12}{4}{1};
    \rectangle{13}{3}{1};
    \rectangle{0}{6}{0};
    \rectangle{1}{5}{0};
    \rectangle{2}{6}{0};
    \rectangle{3}{5}{0};
    \rectangle{4}{6}{0};
    \rectangle{5}{5}{0};
    \rectangle{6}{6}{0};
    \rectangle{7}{5}{0};
    \rectangle{8}{6}{1};
    \rectangle{9}{5}{1};
    \rectangle{10}{6}{1};
    \rectangle{11}{5}{1};
    \rectangle{12}{6}{0};
    \rectangle{13}{5}{0};
    \rectangle{0}{8}{0};
    \rectangle{1}{7}{0};
    \rectangle{2}{8}{0};
    \rectangle{3}{7}{0};
    \rectangle{4}{8}{0};
    \rectangle{5}{7}{0};
    \rectangle{6}{8}{1};
    \rectangle{7}{7}{1};
    \rectangle{8}{8}{0};
    \rectangle{9}{7}{0};
    \rectangle{10}{8}{1};
    \rectangle{11}{7}{0};
    \rectangle{12}{8}{0};
    \rectangle{13}{7}{0};
    \rectangle{0}{10}{0};
    \rectangle{1}{9}{0};
    \rectangle{2}{10}{0};
    \rectangle{3}{9}{0};
    \rectangle{4}{10}{1};
    \rectangle{5}{9}{1};
    \rectangle{6}{10}{0};
    \rectangle{7}{9}{0};
    \rectangle{8}{10}{1};
    \rectangle{9}{9}{1};
    \rectangle{10}{10}{0};
    \rectangle{11}{9}{1};
    \rectangle{12}{10}{1};
    \rectangle{13}{9}{0};
    \rectangle{0}{12}{1};
    \rectangle{1}{11}{0};
    \rectangle{2}{12}{1};
    \rectangle{3}{11}{1};
    \rectangle{4}{12}{0};
    \rectangle{5}{11}{0};
    \rectangle{6}{12}{1};
    \rectangle{7}{11}{1};
    \rectangle{8}{12}{0};
    \rectangle{9}{11}{0};
    \rectangle{10}{12}{0};
    \rectangle{11}{11}{0};
    \rectangle{12}{12}{0};
    \rectangle{13}{11}{1};
    \rectangle{0}{14}{0};
    \rectangle{1}{13}{1};
    \rectangle{2}{14}{0};
    \rectangle{3}{13}{0};
    \rectangle{4}{14}{1};
    \rectangle{5}{13}{1};
    \rectangle{6}{14}{0};
    \rectangle{7}{13}{0};
    \rectangle{8}{14}{0};
    \rectangle{9}{13}{0};
    \rectangle{10}{14}{0};
    \rectangle{11}{13}{0};
    \rectangle{12}{14}{0};
    \rectangle{13}{13}{0};
    \rectangle{0}{16}{1};
    \rectangle{1}{15}{1};
    \rectangle{2}{16}{1};
    \rectangle{3}{15}{1};
    \rectangle{4}{16}{0};
    \rectangle{5}{15}{0};
    \rectangle{6}{16}{0};
    \rectangle{7}{15}{0};
    \rectangle{8}{16}{0};
    \rectangle{9}{15}{0};
    \rectangle{10}{16}{0};
    \rectangle{11}{15}{0};
    \rectangle{12}{16}{0};
    \rectangle{13}{15}{0};
  \end{tikzpicture}
  \caption{\label{fig:dynExample}
    An example of the time evolution. We start with a configuration of $14$ sites
  and evolve it according to the rules~\eqref{eq:rcarules} with periodic
boundary conditions. The black sites behave as particles that move with
velocity $1$ either to the left or to the right. The particles interact
\emph{pairwise}, by annihilating and then reappearing in the next time step
(described by the 3rd, 6th and 8th diagram in~\eqref{eq:rcarules}). This
induces a delay for one site with respect to the original trajectories of the
particles.}
\end{figure}

The model is a~deterministic cellular automaton defined on a~chain of even
length $2n$, with every site being either occupied or empty.  The configuration
of the system at time~$t$ is given by the string of bits
$(s_1^t,s_2^t,\ldots,s_{2n}^t)$, where $s^t_j=0$ if the site $j$ is empty at
time $t$ and $s^t_j=1$ otherwise. The time evolution consists of two time
steps, in the first one only the values on even sites are changed, while
in the second one the odd sites are updated,
\begin{eqnarray}\label{eq:timeProp}
  s_x^{t+1}=\begin{cases}
    \chi(s_{x-1}^t,s_x^t,s_{x+1}^t);\qquad & x+t\equiv 0\pmod{2},\\
    s_x^t;& x+t\equiv 1\pmod{2},
  \end{cases}
\end{eqnarray}
where we assume periodic boundaries, $s^t_{2n+k}\equiv s^t_{k}$.
The local update map $\chi$ is given by the \emph{Rule 54} (RCA54) of Bobenko~\emph{et.\ al.},
as introduced in~\cite{bobenko1993two},
\begin{eqnarray}
  \chi(s_1,s_2,s_3)=(s_1+s_2+s_3+s_1 s_3)\pmod{2}.
\end{eqnarray}
It is convenient to imagine the chain to have a~zig-zag shape, as schematically shown
in Figure~\ref{fig:lattGeo}.
Then, the local update rule can be expressed graphically by representing
occupied sites by black and empty sites by white squares,
\begin{eqnarray}\label{eq:rcarules}\fl
  \rcaTextRule{0}{0}{0}{0}\quad
  %\rcaRule{0}{0}{0}{0}\quad
  \rcaRule{0}{0}{1}{1}\quad
  \rcaRule{0}{1}{0}{1}\quad
  \rcaRule{0}{1}{1}{0}\quad
  \rcaRule{1}{0}{0}{1}\quad
  \rcaRule{1}{0}{1}{1}\quad
  \rcaRule{1}{1}{0}{0}\quad
  \rcaRule{1}{1}{1}{0}\enskip
  ,
\end{eqnarray}
where the bottom three squares represent a configuration of three consecutive
sites at time $t$, $(s_1,s_2,s_3)$, while the top square is the updated site,
$s_2^{\prime}=\chi(s_1,s_2,s_3)$.  An example of the time evolution induced by
these rules is shown in Figure~\ref{fig:dynExample}.
The graphical representation provides an intuitive interpretation of the
dynamics; occupied sites behave as particles that move with velocity $\pm 1$
while undisturbed. When two oppositely moving particles meet, they interact
by being delayed for one site (or one time step) with respect to their initial trajectories.

\subsection{Time evolution of probability distributions}
\emph{The states} $\vec{p}$ are probability distributions over configurations
and can be expressed as vectors in $\mathbb{R}^{2^{2n}}$,
with nonnegative components and the appropriate normalization, 
\begin{eqnarray}
  \vec{p}=(p_0,p_1,\ldots,p_{2^{2n}-1}),\qquad p_s\ge0,\qquad \sum_{s=0}^{2^{2n}-1}p_s=1,
\end{eqnarray}
where $p_s$ is the probability of observing the~configuration
$(s_1,s_2,\ldots,s_{2n})$, given by the binary representation of $s$:
$s=\sum_{j=1}^{2n} 2^{2n-,j} s_j$. The local update rule is given in terms
of the three-site permutation matrix~$P$ with the elements
$P_{(s_1 s_2 s_3),(b_1 b_2 b_3)} = \delta_{s_1,b_1} \delta_{s_2,\chi(b_1,b_2,b_3)} \delta_{s_3,b_3}$.
Explicitly,
\begin{eqnarray}
  P=\begin{bmatrix}[0.8]
    1 & & & & & & & \\
    & & & 1 & & & & \\
    & & 1 & & & & & \\
    & 1 & & & & & & \\
    & & & & & & 1 & \\
    & & & & & & & 1 \\
    & & & & 1 & & & \\
    & & & & & 1 & & \\
  \end{bmatrix},\qquad P^2=\unit.
\end{eqnarray}
Introducing the local propagators of the triplets of neighbouring sites as
$P_{x-1,x,x+1}=\unit_{2^{x-2}}\otimes P \otimes \unit_{2^{2n-x-1}}$,
the time evolution of the state $\vec{p}^t$ can be expressed in the following way,
\begin{eqnarray}
  \eqalign{
    \vec{p}^{t+1}=\begin{cases}
      U_{\mathrm{e}} \vec{p}^t;& t\equiv 0\pmod{2},\\
      U_{\mathrm{o}} \vec{p}^t;& t\equiv 1\pmod{2},
    \end{cases}\\
    U_{\mathrm{e}}=\prod_{x=1}^n P_{2x-1,2x,2x+1},\qquad
    U_{\mathrm{o}}=\prod_{x=1}^n P_{2x-2,2x-1,2x}.
  }
\end{eqnarray}

\subsection{Equilibrium states}
The equilibrium states should be invariant under the time evolution for whole
periods of time, i.e.\ for even $t$. Since the time propagation is staggered,
we can introduce two versions of the equilibrium state, $\vec{p}$ and
$\vec{p}^{\prime}$, corresponding to the even and the odd time steps. In this case the
time-invariance condition takes the following form,
\begin{eqnarray}
  \vec{p}^{\prime}=U_{\mathrm{e}}\vec{p},\qquad
  \vec{p}=U_{\mathrm{o}}\vec{p}^{\prime}.
\end{eqnarray}
In the paper we consider the states that can be expressed in a matrix-product form
similar to the stationary states introduced in~\cite{prosenBucaCA54,bucaetalLargeDev}.
Let $\vec{W}(\xi,\omega)$ and $\vec{W}^{\prime}(\xi,\omega)$ be vectors in the physical space,
\begin{eqnarray}
  \vec{W}(\xi,\omega)=\begin{bmatrix}W_0(\xi,\omega)\\W_1(\xi,\omega)\end{bmatrix},\qquad
  \vec{W}^{\prime}(\xi,\omega)=\begin{bmatrix}W_0^{\prime}(\xi,\omega)\\W_1^{\prime}(\xi,\omega)\end{bmatrix}.
\end{eqnarray}
Their components are $3\times 3$ matrices that depend on two \emph{spectral parameters} $\xi$ and $\omega$;
$\xi,\omega>0$,
\begin{eqnarray}\label{eq:matricesWWp}\fl
    W_0(\xi,\omega)=\begin{bmatrix}1& 0& 0\\ \xi& 0& 0\\ 1& 0& 0\end{bmatrix}=W_0^{\prime}(\omega,\xi),\qquad
    W_1(\xi,\omega)=\begin{bmatrix}0& \xi& 0\\ 0& 0& 1\\ 0& 0& \omega\end{bmatrix}=W_1^{\prime}(\omega,\xi).
\end{eqnarray}
The two sets of matrices $\vec{W}^{\prime}(\xi,\omega)$,
$\vec{W}^{\prime}(\xi,\omega)$ are mapped to each other under the exchange of
the (spectral) parameters $\xi\leftrightarrow\omega$. In what follows, the explicit
dependence on $\xi$ and $\omega$ will sometimes be suppressed to simplify the notation.
The matrices obey a~cubic algebraic relation
\begin{eqnarray}\label{eq:cubicRelation}
  P_{123}\vec{W}_1\vec{W}^{\prime}_2\vec{W}_3 S=\vec{W}_1 S \vec{W}_2 \vec{W}_3^{\prime},
\end{eqnarray}
where $S$ is a~$3\times 3$ matrix acting in the auxiliary space:
\begin{eqnarray}
  S=\begin{bmatrix}
    1 & & \\
    & & 1 \\
    & 1 &
  \end{bmatrix},\qquad
  S^2=\unit.
\end{eqnarray}
The relation, spelled out in physical components, explicitly states
$W_{s_1} W^{\prime}_{\chi(s_1,s_2,s_3)} W_{s_3} S= W_{s_1} S W_{s_2} W^{\prime}_{s_3}$,
for any combination of $s_1,s_2,s_3\in\{0,1\}$.
Due to the mapping between $\vec{W}$
and $\vec{W}^{\prime}$, the following dual relation also holds,
\begin{eqnarray}\label{eq:dualCubicRelation}
P_{123}\vec{W}^{\prime}_1\vec{W}_2\vec{W}^{\prime}_3 S= \vec{W}^{\prime}_1 S \vec{W}^{\prime}_2 \vec{W}_3.
\end{eqnarray}
The equilibrium states can then be expressed as
\begin{eqnarray}
   \label{eq:invst}
    \vec{p}^{(2n)}=\frac{1}{Z}\tr\big(\vec{W}_1\vec{W}_2^{\prime}\cdots\vec{W}_{2n}^{\prime}\big),\qquad
    \vec{p}^{\prime(2n)}=\frac{1}{Z}\tr\big(\vec{W}_1^{\prime}\vec{W}_2\cdots\vec{W}_{2n}\big),
\end{eqnarray}
where $Z$ is fixed by the normalization. Using the relations~\eqref{eq:cubicRelation}
and~\eqref{eq:dualCubicRelation}, together with properties $S^2=\unit$ and $P^2=\unit$,
it is straightforward to see that the states $\vec{p}^{(2n)}$ and $\vec{p}^{\prime(2n)}$
are mapped to each other under the time propagation,
\begin{eqnarray}
  \vec{p}^{(2n)}=U_{\mathrm{o}} \vec{p}^{\prime(2n)},\qquad 
  \vec{p}^{\prime(2n)}=U_{\mathrm{e}} \vec{p}^{(2n)},
\end{eqnarray}
which implies their stationarity. Note that the exchange of the parameters
$\xi\leftrightarrow\omega$ shifts the state for one time step. Furthermore, the
exchange of parameters also corresponds to the shift for one lattice site.
Explicitly,
\begin{eqnarray}\label{eq:transSymmetry}
  \eta\left(\vec{p}^{(2n)}\right) = \vec{p}^{\prime(2n)},\qquad
  \eta\left(\vec{p}^{\prime(2n)}\right) = \vec{p}^{(2n)},
\end{eqnarray}
where $\eta$ is the one-lattice-site shift operator, given in terms of a $2^{2n}\times 2^{2n}$ matrix
$\eta_{(s_1 s_2 \ldots s_{2n}),(b_1 b_2 \ldots b_{2n})} = 
\delta_{s_2,b_1}\delta_{s_3,b_2}\delta_{s_4,b_3}\cdots
\delta_{s_{2n},b_{2n-1}}\delta_{s_1,b_{2n}}$.
It can be argued that \eqref{eq:invst} provides a two-parameter
$(\xi,\omega\in\RR_{\ge 0})$ family of finitely correlated equilibrium states
of the model, a kind of a caricature of generalized Gibbs states.

\subsection{Asymptotic equilibrium distributions of shorter configurations}
The equilibrium distribution on a~smaller subchain of length $2\, m$, $m\le n$,
can be expressed by summing over all the other sites
\begin{eqnarray}
  \vec{p}_{[1,2m]}^{(2n)}=\frac{\tr\big(
  {\vec{W}_1\vec{W}_2^{\prime}\vec{W}_{3}\cdots \vec{W}^{\prime}_{2m}}T^{n-m}\big)}{\tr T^n},
\end{eqnarray}
with the transfer matrix $T=(W_0+W_1)(W_0^{\prime}+W_1^{\prime})$ taking the form
\begin{eqnarray}
  T=\begin{bmatrix}
    1+\xi\omega & \omega & \xi \\
    1+\xi & \xi\omega & \xi \\
    1+\omega & \omega & \xi \omega
  \end{bmatrix}.
\end{eqnarray}
Fixing the subchain length $m$ while taking the thermodynamic limit $n\to\infty$ this reduces to
\begin{eqnarray}\label{eq:finiteProbDistr}
  \vec{p}_{[1,2m]}
  =\lim_{n\to\infty}\vec{p}_{[1,2m]}^{(2n)}=\lambda^{-m}
\frac{\bra{l}\vec{W}_1\vec{W}_2^{\prime}\vec{W}_3\cdots\vec{W}^{\prime}_{2m}\ket{r}}{\braket{l}{r}},
\end{eqnarray}
where $\lambda$ is the leading eigenvalue of $T$,
\begin{eqnarray}\label{eq:defLambda}\fl
  \eqalign{
    \lambda(\xi,\omega)=\frac{1}{3}\left(1+3 \xi\omega
    +\left(\frac{\Delta}{2}\right)^{1/3}\!\!+(1+3\xi)(1+3\omega)\left(\frac{2}{\Delta}\right)^{1/3}\right),\\
    \Delta=2+9(\xi+\omega)+27 \xi \omega(2+\xi+\omega)+\sqrt{27(\xi-\omega)^2(9\xi\omega(3\xi\omega-2)-4(\xi+\omega)-1)},
    }
  \end{eqnarray}
and $\bra{l}$, $\ket{r}$ are the corresponding left and right (unnormalized) eigenvectors,
\begin{eqnarray}
  \eqalign{
    \ket{l}=\begin{bmatrix}
      (\lambda-\xi\omega)^2-\xi\omega\\
      \omega(\lambda-\xi\omega+\xi)\\
      \xi(\lambda-\xi\omega+\omega)
    \end{bmatrix},\qquad
    \ket{r}=\begin{bmatrix}
      \omega(\lambda-\xi\omega+\xi)\\
      (\lambda-\xi\omega)^2-\lambda-\xi\\
      \omega(\lambda-\xi\omega+\omega)
    \end{bmatrix}.
  }
\end{eqnarray}
Note that the leading eigenvalue $\lambda(\xi,\omega)$ is the largest solution
of the cubic equation
\begin{eqnarray}
  \lambda^3-\lambda^2(1+3\xi\omega)-\lambda(\xi+\omega+\xi\omega(1-3\xi\omega))-\xi\omega(1-\xi\omega)^2=0.
\end{eqnarray}
It is larger than $1$ and appears as a single (isolated) real root for all
non-negative values of the parameters $\xi$, $\omega$.

The expression~\eqref{eq:finiteProbDistr} holds for all finite subsections of the
chain that start at odd sites at even times or even sites at odd times,
i.e.\ the components of $\vec{p}$ are probabilities of observing configurations
$(s_x^t,s_{x+1}^t,\ldots,s_{x+2m-1}^t)$ if $x+t\equiv 1\pmod{2}$
holds.~\footnote{
  Since $\vec{p}_{[1,2m]}$ is well defined for any~$m$, the
  subscript will be omitted and the exact length specified when ambiguous. In what follows,
$\vec{p}$ will always refer to the asymptotic equilibrium distribution.
}
In other case, we have to exchange the role of the parameters $\xi$ and $\omega$. This
can be summarized as
\begin{eqnarray}\label{eq:asymptoticDistrs}
  \eqalign{
    p\Big(\begin{tikzpicture}[baseline={([yshift=-0.5ex]current bounding box.center)},scale=1.5] 
        \textrectangle{0}{0.5}{$s_1$};
        \textrectangle{1}{-0.5}{$s_2$};
        \textrectangle{2}{0.5}{$\cdots$};
        \textrectangle{3}{-0.5}{};
        \textrectangle{4}{0.5}{};
        \textrectangle{5}{-0.5}{};
        \textrectangle{6}{0.5}{$\cdots$};
        \textrectangle{7}{-0.5}{$s_{2m}$};
    \end{tikzpicture}\Big) \coloneqq
    p_{s_1 s_2\ldots s_{2m}}:=\frac{\lambda^{-m}}{\braket{l}{r}}\bra{l}W_{s_1}W^{\prime}_{s_2}\cdots W^{\prime}_{s_{2m}}\ket{r},\\
    p\Big(\begin{tikzpicture}[baseline={([yshift=-0.5ex]current bounding box.center)},scale=1.5] 
        \textrectangle{0}{-0.5}{$s_1$};
        \textrectangle{1}{0.5}{$s_2$};
        \textrectangle{2}{-0.5}{$\cdots$};
        \textrectangle{3}{0.5}{};
        \textrectangle{4}{-0.5}{};
        \textrectangle{5}{0.5}{};
        \textrectangle{6}{-0.5}{$\cdots$};
        \textrectangle{7}{0.5}{$s_{2m}$};
    \end{tikzpicture}\Big)\coloneqq
    p^{\prime}_{s_1 s_2\ldots s_{2m}}
    :=\frac{\lambda^{-m}}{\braket{l^{\prime}}{r^{\prime}}}\bra{l^{\prime}}W^{\prime}_{s_1}W_{s_2}\cdots W_{s_{2m}}\ket{r^{\prime}},
  }
\end{eqnarray}
where the new (primed) left/right vectors are obtained from the old (unprimed) ones by exchanging $\xi\leftrightarrow\omega$:
$\bra{l^{\prime}(\xi,\omega)}=\bra{l(\omega,\xi)}$ and $\ket{r^{\prime}(\xi,\omega)}=\ket{r(\omega,\xi)}$.

The asymptotic equilibrium distribution on~$2m$ sites~\eqref{eq:finiteProbDistr}
can be alternatively understood as a~steady-state solution to a~specific boundary
driven setup, where the bulk dynamics is given by the deterministic update
rules~\eqref{eq:rcarules}, while the boundary sites change stochastically
(i.e.\ the situation studied
in~\cite{prosenMejiaMonasterioCA54,inoueTakesueCA54,prosenBucaCA54,bucaetalLargeDev}).
In~\ref{app:boundaryProps} we provide the details about the equivalent
stochastic-boundary setup.

\subsection{Statistical independence of solitons}
The stationary (equilibrium) distributions~$\vec{p}(\xi,\omega)$,
Eq.~\eqref{eq:invst}, are characterized by \emph{statistical independence
of solitons}, i.e.\ the probability of encountering
a~soliton is the same everywhere, independently of the positions of other
solitons. This is a consequence of the fact that the conditional probability of observing
$(s_{2k-1},s_{2k})$, given the previous configuration
$(s_1,s_2,\ldots,s_{2k-2})$, depends only on the value of the last four bits,
$(s_{2k-3},s_{2k-2},s_{2k-1}s_{2k})$.  Explicitly,
\begin{eqnarray}\label{eq:cond4site1}\fl
  \frac{p_{s_1s_2\ldots s_{2k-1}s_{2k}}}{p_{s_1 s_2\ldots s_{2k-2}}}=
  \frac{p_{s_{2k-3}s_{2k-2}s_{2k-1}s_{2k}}}{p_{s_{2k-3} s_{2k-2}}},\qquad
  \frac{p^{\prime}_{s_1s_2\ldots s_{2k-1}s_{2k}}}{p^{\prime}_{s_1 s_2\ldots s_{2k-2}}}=
  \frac{p^{\prime}_{s_{2k-3}s_{2k-2}s_{2k-1}s_{2k}}}{p^{\prime}_{s_{2k-3} s_{2k-2}}}.
\end{eqnarray}
A~similar relation holds for the conditional probability of finding $(s_1,s_2)$ on the first
two sites given that the sites from $3$ to $2k$ are in the configuration
$(s_3,s_4,\ldots s_{2k})$;
\begin{eqnarray}\label{eq:cond4site2}
  \frac{p_{s_1 s_2\ldots s_{2k}}}{p_{s_3 s_4\ldots s_{2k}}}
  =\frac{p_{s_1 s_2 s_3 s_4}}{p_{s_3 s_4}},\qquad
  \frac{p^{\prime}_{s_1 s_2\ldots s_{2k}}}{p^{\prime}_{s_3 s_4\ldots s_{2k}}}
  =\frac{p^{\prime}_{s_1 s_2 s_3 s_4}}{p^{\prime}_{s_3 s_4}}.
\end{eqnarray}
A proof of these relations simply follows from the sparse structure
of the matrices $W_s$, $W^{\prime}_s$, as is shown in~\ref{app:condProbs}.
Physically, this means that due to the pure \emph{contact}  (ultra local)
interactions between particles the equilibrium state can be interpreted as an ideal
gas of solitons. 

Taking into account~\eqref{eq:cond4site1} and~\eqref{eq:cond4site2} we can now
introduce probabilities of observing the left/right movers, $(p_l,p_r)$.  The first
parameter, $p_l$, is the conditional probability of observing a~left mover, if
we know that the neighbouring left ray does not contain a~left moving soliton.
This can be easily expressed in terms of equilibrium probabilities as
\begin{eqnarray}\label{eq:probLeft}
  p_l=\frac{p\big(\!\leftSaw{0}{0}{1}{0}\!\big)+p\big(\!\leftSaw{0}{0}{1}{1}\!\big)}{p\big(\!\leftsSaw{0}{0}\!\big)}
  =\frac{p\big(\!\leftSaw{0}{1}{0}{0}\!\big)+p\big(\!\leftSaw{0}{1}{0}{1}\!\big)}{p\big(\!\leftsSaw{0}{1}\!\big)}
  =\frac{\xi(\lambda+\omega(1-\xi))}{\lambda(1+\xi)+\xi(1-\xi\omega)}.
\end{eqnarray}
Similarly, $p_r$ is the conditional probability of observing a right moving soliton,
provided that the neighbouring right ray does not contain a right mover,
\begin{eqnarray}\label{eq:probRight}
  p_r=\frac{p\big(\!\rightSaw{0}{1}{0}{0}\!\big)+p\big(\!\rightSaw{1}{1}{0}{0}\!\big)}{p\big(\!\rightsSaw{0}{0}\!\big)}
  =\frac{p\big(\!\rightSaw{0}{0}{1}{0}\!\big)+p\big(\!\rightSaw{1}{0}{1}{0}\!\big)}{p\big(\!\rightsSaw{1}{0}\!\big)}
  =\frac{\omega(\lambda+\xi(1-\omega))}{\lambda(1+\omega)+\omega(1-\xi\omega)}.
\end{eqnarray}
Note that the shift for one time step or one lattice site maps $p_r$ to $p_l$ and vice versa;
$p_l \xleftrightarrow{\xi \leftrightarrow \omega} p_r$.

\section{Probability distributions of configurations in time}\label{sec:vState}
We now proceed to the main result of the paper, namely the construction of the matrix
product representation of the~\emph{time state} $\vec{q}$. The components of
the probability vector~$q_{b_0 b_1\ldots b_{m-1}}$ are the probabilities of
encountering multi-time configurations encoded by bit strings $(b_0,b_1,\ldots
b_{m-1})\in\{0,1\}^m$ in the middle vertical saw, while the horizontal chain
configurations (say along the bottom saw) are distributed according to the
equilibrium state~$\vec{p}$ (schematically depicted in
Fig.~\ref{fig:schPicture}). Explicitly,
\begin{eqnarray}\label{eq:eqToSolve}
q_{b_0\,b_1\,\ldots b_{m-1}}=\smashoperator{\sum_{\underline{s}\equiv(s_{-m},s_{-m+1},\ldots,s_{m-1})}}p_{\underline{s}}\,
\delta_{b_0,s^0_{0}}
  \delta_{b_1,s^1_{1}}
  \delta_{b_2,s^2_{0}}
  \delta_{b_3,s^3_{1}}
  \cdots
  \delta_{b_{m-1},s^{m-1}_{{\mathrm{mod}}(m-1,2)}},
\end{eqnarray}
where $s^t_k$ is the value at site $k$ of the configuration
$\underline{s}=(s_{-m},s_{-m+1},\ldots s_{m-1})$ propagated for $t$ time steps,
starting from the initial data $s^0_k = s_k$. Due to the staggering, we should
consider the site $0$ at even time steps and the site $1$ at odd time steps.
The last site, whether $0$ or $1$, therefore depends on the parity of
$m$.~\footnote{Note also that the labels of the sites now go from~$-m$ to $m-1$
  and no longer from $1$ to $2m$.  The convention is such, that the sites with the
even labels are getting propagated in the first half step and the sites with
the odd labels in the second one, irrespective of the parity of $m$.}

{
  The definition of the time-state components~\eqref{eq:eqToSolve} does not
  explicitly rely on the particularities of the model and can be
  directly applied to any system, for which the time-evolved
  configurations~$\underline{s}^t$ are in one-to-one correspondence with the
  initial configurations~$\underline{s}^0$, i.e.\ any deterministic classical
  system with a finite configuration space. Moreover, the time-state
  can formally be defined for any underlying statistical state by replacing the
  probabilities~$p_{\underline{s}}$ accordingly. However, as we show below, for
  our specific choice of asymptotic stationary probability
  distribution~\eqref{eq:asymptoticDistrs} the time-state greatly simplifies,
  which is not true in general.
}

Since the positions of the solitons in the equilibrium state $\vec{p}$
are statistically independent, the probability of a~left (right) mover
reaching the central site at time $t$ does not depend on the
previously observed solitons, as long as there was no left (right)
mover in the previous time step $t-1$. Therefore, the conditional
probability of observing a~vertical configuration $(b_0, b_1, b_2,\ldots b_{k-1}, b_k)$
given the previous configuration $(b_0, b_1, b_2, \ldots b_{k-1})$,
depends only on the last $4$ bits. This implies a~relation
similar to~\eqref{eq:cond4site1} and~\eqref{eq:cond4site2},
\begin{eqnarray}\fl
  \frac{q_{b_0 b_1\ldots b_{2 k-1}}}{q_{b_0 b_1 \ldots b_{2 k-2}}}=
  f(b_{2k-4},b_{2k-3},b_{2k-2},b_{2k-1}),\quad
  \frac{q_{b_0 b_1\ldots b_{2 k}}}{q_{b_0 b_1 \ldots b_{2 k-1}}}=
  f^{\prime}(b_{2k-3},b_{2k-2},b_{2k-1},b_{2k}),
\end{eqnarray}
where we introduced two yet unknown functions of the last four bits,
$f,f^{\prime}: \mathbb{Z}_2\times \mathbb{Z}_2\times \mathbb{Z}_2\times \mathbb{Z}_2\to \mathbb{R}_{\ge 0}$,
that have to be mapped into each other under the exchange of the spectral parameters
$\xi$ and $\omega$ (or equivalently, $p_r$ and $p_l$),
\begin{eqnarray}
  f(b_{1},b_{2},b_{3},b_{4})\xleftrightarrow{p_l\leftrightarrow p_r}
  f^{\prime}(b_{1},b_{2},b_{3},b_{4}).
\end{eqnarray}
To determine the functions $f$, $f^{\prime}$, one should first classify all
the vertical configurations of $3$ sites and their
transitions into $4$-site configurations. They can be divided into
$3$ different types:
\begin{itemize}
  \item A configuration might be \emph{inaccessible}, i.e.\ the configuration
    is inconsistent with the time evolution rules~\eqref{eq:rcarules}. The
    probability of obtaining such configurations is $0$. Such are the
    following configurations:
    \begin{eqnarray}\label{eq:conf1}
      \begin{tikzpicture}[baseline={(current bounding box.center)}] 
        \configThreeLeft{0}{1}{0};
      \end{tikzpicture}\qquad
      \begin{tikzpicture}[baseline={(current bounding box.center)}] 
        \configThreeLeft{1}{1}{1};
      \end{tikzpicture}\qquad
      \begin{tikzpicture}[baseline={(current bounding box.center)}] 
        \configThreeRight{0}{1}{0};
      \end{tikzpicture}\qquad
      \begin{tikzpicture}[baseline={(current bounding box.center)}] 
        \configThreeRight{1}{1}{1};
      \end{tikzpicture}.
    \end{eqnarray}
  \item A configuration is allowed and the local update rules deterministically determine the next bit,
    i.e.\ the conditional probability $q_{b_{k-3}b_{k-2}b_{k-1}b_{k}}/q_{b_{k-3}b_{k-2}b_{k-1}}$
    of observing $b_{k}$ is either $0$ or $1$. The following $3$ configurations are of this type:
    \begin{eqnarray}\label{eq:conf2l}
        \begin{tikzpicture}[baseline={(current bounding box.center)}] 
          \node (A) at (0,0){
            \begin{tikzpicture}
              \configThreeRight{0}{0}{1};
            \end{tikzpicture}
          };
          \node (B) at (1.75,0){
            \begin{tikzpicture}
              \configFourRight{0}{0}{1}{1};
            \end{tikzpicture}
          };
          \draw [->] (A) -- (B) node[midway,above] {$\scriptstyle 1$};
        \end{tikzpicture}\qquad
        \begin{tikzpicture}[baseline={(current bounding box.center)}] 
          \node (A) at (0,0){
            \begin{tikzpicture}
              \configThreeRight{1}{0}{1};
            \end{tikzpicture}
          };
          \node (B) at (1.75,0){
            \begin{tikzpicture}
              \configFourRight{1}{0}{1}{1};
            \end{tikzpicture}
          };
          \draw [->] (A) -- (B) node[midway,above] {$\scriptstyle 1$};
        \end{tikzpicture}\qquad
        \begin{tikzpicture}[baseline={(current bounding box.center)}] 
          \node (A) at (0,0){
            \begin{tikzpicture}
              \configThreeRight{0}{1}{1};
            \end{tikzpicture}
          };
          \node (B) at (1.75,0){
            \begin{tikzpicture}
              \configFourRight{0}{1}{1}{0};
            \end{tikzpicture}
          };
          \draw [->] (A) -- (B) node[midway,above] {$\scriptstyle1$};
        \end{tikzpicture},
      \end{eqnarray}
      and  analogously for the other parity of $k$ (i.e.\ considering the flipped configurations),
    \begin{eqnarray}\label{eq:conf2r}
        \begin{tikzpicture}[baseline={(current bounding box.center)}] 
          \node (A) at (0,0){
            \begin{tikzpicture}
              \configThreeLeft{0}{0}{1};
            \end{tikzpicture}
          };
          \node (B) at (1.75,0){
            \begin{tikzpicture}
              \configFourLeft{0}{0}{1}{1};
            \end{tikzpicture}
          };
          \draw [->] (A) -- (B) node[midway,above] {$\scriptstyle 1$};
        \end{tikzpicture}\qquad
        \begin{tikzpicture}[baseline={(current bounding box.center)}] 
          \node (A) at (0,0){
            \begin{tikzpicture}
              \configThreeLeft{1}{0}{1};
            \end{tikzpicture}
          };
          \node (B) at (1.75,0){
            \begin{tikzpicture}
              \configFourLeft{1}{0}{1}{1};
            \end{tikzpicture}
          };
          \draw [->] (A) -- (B) node[midway,above] {$\scriptstyle 1$};
        \end{tikzpicture}\qquad
        \begin{tikzpicture}[baseline={(current bounding box.center)}] 
          \node (A) at (0,0){
            \begin{tikzpicture}
              \configThreeLeft{0}{1}{1};
            \end{tikzpicture}
          };
          \node (B) at (1.75,0){
            \begin{tikzpicture}
              \configFourLeft{0}{1}{1}{0};
            \end{tikzpicture}
          };
          \draw [->] (A) -- (B) node[midway,above] {$\scriptstyle1$};
        \end{tikzpicture}.
      \end{eqnarray}
      Complementary diagrams with negated $b_k$ are associated with conditional probability $0$.
    \item A configuration is allowed and both the configurations $(b_{k-3} b_{k-2} b_{k-1} 0)$
      and $(b_{k-3} b_{k-2} b_{k-1} 1)$ are also allowed. These are the following:
    \begin{eqnarray}\label{eq:conf3l}
      \begin{tikzpicture}[baseline={(current bounding box.center)}] 
        \node (A) at (0,0){
          \begin{tikzpicture}
            \configThreeRight{0}{0}{0};
          \end{tikzpicture}
        };
        \node (B) at (1.75,0.7){
          \begin{tikzpicture}
            \configFourRight{0}{0}{0};
            \draw [->,>=stealth,red,thick] ({2.5*\a-0.25*\a},{0+0.25*\a}) -- ({1.5*\a-0.25*\a},{\a+0.25*\a});
          \end{tikzpicture}
        };
        \node (C) at (1.75,-0.7){
          \begin{tikzpicture}
            \configFourRight{0}{0}{0}{1};
            \draw [->,>=stealth,red,thick] ({2.5*\a-0.25*\a},{0+0.25*\a}) -- ({1.5*\a-0.25*\a},{\a+0.25*\a});
          \end{tikzpicture}
        };
        \draw [->] (A) -- (B) node [midway, above] {$\scriptstyle{1-p_l}$};
        \draw [->] (A) -- (C) node [midway, above] {$\scriptstyle p_l$};
      \end{tikzpicture}\qquad
      \begin{tikzpicture}[baseline={(current bounding box.center)}] 
        \node (A) at (0,0){
          \begin{tikzpicture}
            \configThreeRight{1}{0}{0};
          \end{tikzpicture}
        };
        \node (B) at (1.75,0.7){
          \begin{tikzpicture}
            \configFourRight{1}{0}{0}{0};
            \draw [->,>=stealth,red,thick] ({2.5*\a-0.25*\a},{0+0.25*\a}) -- ({1.5*\a-0.25*\a},{\a+0.25*\a});
          \end{tikzpicture}
        };
        \node (C) at (1.75,-0.7){
          \begin{tikzpicture}
            \configFourRight{1}{0}{0}{1};
            \draw [->,>=stealth,red,thick] ({2.5*\a-0.25*\a},{0+0.25*\a}) -- ({1.5*\a-0.25*\a},{\a+0.25*\a});
          \end{tikzpicture}
        };
        \draw [->] (A) -- (B) node [midway, above] {$\scriptstyle{1-p_l}$};
        \draw [->] (A) -- (C) node [midway, above] {$\scriptstyle p_l$};
      \end{tikzpicture}
      \qquad
      \begin{tikzpicture}[baseline={(current bounding box.center)}] 
        \node (A) at (0,0){
          \begin{tikzpicture}
            \configThreeRight{1}{1}{0};
          \end{tikzpicture}
        };
        \node (B) at (1.75,0.7){
          \begin{tikzpicture}
            \configFourRight{1}{1}{0}{0};
            \draw [->,>=stealth,red,thick] ({2.5*\a-0.25*\a},{-2*\a+0.25*\a}) -- ({1.5*\a-0.25*\a},{-\a+0.25*\a});
          \end{tikzpicture}
        };
        \node (C) at (1.75,-0.7){
          \begin{tikzpicture}
            \configFourRight{1}{1}{0}{1};
            \draw [->,>=stealth,red,thick] ({2.5*\a-0.25*\a},{-2*\a+0.25*\a}) -- ({1.5*\a-0.25*\a},{-\a+0.25*\a});
          \end{tikzpicture}
        };
        \draw [->] (A) -- (B) node [midway, above] {$\scriptstyle{1-p_l}$};
        \draw [->] (A) -- (C) node [midway, above] {$\scriptstyle p_l$};
      \end{tikzpicture}
      .
    \end{eqnarray}
        The conditional probabilities $q_{b_{k-3}b_{k-2}b_{k-1}b_{k}}/q_{b_{k-3}b_{k-2}b_{k-1}}$
    correspond to the probability of a~left mover (not)reaching the vertical saw
    (corresponding to the top/bottom option respectively) at the position indicated
    by the red arrows. Since the solitons in the equilibrium state are independent,
    the probability~$p_l$ is given by~\eqref{eq:probLeft}. For the flipped configurations,
    everything works similarly, only $p_l$ has to be replaced with $p_r$ from~\eqref{eq:probRight},
    \begin{eqnarray}\label{eq:conf3r}
      \begin{tikzpicture}[baseline={(current bounding box.center)}] 
        \node (A) at (0,0){
          \begin{tikzpicture}
            \configThreeLeft{0}{0}{0};
          \end{tikzpicture}
        };
        \node (B) at (1.75,0.7){
          \begin{tikzpicture}
            \configFourLeft{0}{0}{0};
            \draw [->,>=stealth,red,thick] ({-2.5*\a+0.25*\a},{0+0.25*\a}) -- ({-1.5*\a+0.25*\a},{\a+0.25*\a});
          \end{tikzpicture}
        };
        \node (C) at (1.75,-0.7){
          \begin{tikzpicture}
            \configFourLeft{0}{0}{0}{1};
            \draw [->,>=stealth,red,thick] ({-2.5*\a+0.25*\a},{0+0.25*\a}) -- ({-1.5*\a+0.25*\a},{\a+0.25*\a});
          \end{tikzpicture}
        };
        \draw [->] (A) -- (B) node [midway, above] {$\scriptstyle{1-p_r}$};
        \draw [->] (A) -- (C) node [midway, above] {$\scriptstyle p_r$};
      \end{tikzpicture}\qquad
      \begin{tikzpicture}[baseline={(current bounding box.center)}] 
        \node (A) at (0,0){
          \begin{tikzpicture}
            \configThreeLeft{1}{0}{0};
          \end{tikzpicture}
        };
        \node (B) at (1.75,0.7){
          \begin{tikzpicture}
            \configFourLeft{1}{0}{0}{0};
            \draw [->,>=stealth,red,thick] ({-2.5*\a+0.25*\a},{0+0.25*\a}) -- ({-1.5*\a+0.25*\a},{\a+0.25*\a});
          \end{tikzpicture}
        };
        \node (C) at (1.75,-0.7){
          \begin{tikzpicture}
            \configFourLeft{1}{0}{0}{1};
            \draw [->,>=stealth,red,thick] ({-2.5*\a+0.25*\a},{0+0.25*\a}) -- ({-1.5*\a+0.25*\a},{\a+0.25*\a});
          \end{tikzpicture}
        };
        \draw [->] (A) -- (B) node [midway, above] {$\scriptstyle{1-p_r}$};
        \draw [->] (A) -- (C) node [midway, above] {$\scriptstyle p_r$};
      \end{tikzpicture}
      \qquad
      \begin{tikzpicture}[baseline={(current bounding box.center)}] 
        \node (A) at (0,0){
          \begin{tikzpicture}
            \configThreeLeft{1}{1}{0};
          \end{tikzpicture}
        };
        \node (B) at (1.75,0.7){
          \begin{tikzpicture}
            \configFourLeft{1}{1}{0}{0};
            \draw [->,>=stealth,red,thick] ({-2.5*\a+0.25*\a},{-2*\a+0.25*\a}) -- ({-1.5*\a+0.25*\a},{-\a+0.25*\a});
          \end{tikzpicture}
        };
        \node (C) at (1.75,-0.7){
          \begin{tikzpicture}
            \configFourLeft{1}{1}{0}{1};
            \draw [->,>=stealth,red,thick] ({-2.5*\a+0.25*\a},{-2*\a+0.25*\a}) -- ({-1.5*\a+0.25*\a},{-\a+0.25*\a});
          \end{tikzpicture}
        };
        \draw [->] (A) -- (B) node [midway, above] {$\scriptstyle{1-p_r}$};
        \draw [->] (A) -- (C) node [midway, above] {$\scriptstyle p_r$};
      \end{tikzpicture}
      .
    \end{eqnarray}
\end{itemize}
Extracting the precise values of~$f(b_1,b_2,b_3,b_4)$ and $f^{\prime}(b_1,b_2,b_3,b_4)$
from the diagrams~(\ref{eq:conf1}--\ref{eq:conf3r}),
%\eqref{eq:conf1},\eqref{eq:conf2r},\eqref{eq:conf2l},\eqref{eq:conf3l},\eqref{eq:conf3l},\eqref{eq:conf3r}
the probability of a~vertical configuration can be explicitly expressed as
\begin{eqnarray}
\label{eq:tv}
  \fl
  q_{b_0 b_1\ldots b_k}=q_{b_0 b_1 b_2}
  f(b_0,b_1,b_2,b_3)
  f^{\prime}(b_1,b_2,b_3,b_4)
  \cdots
  f^{(\prime)}(b_{k-1},b_{k-2},b_{k-1},b_k),
\end{eqnarray}
where the last value is $f(b_{k-1},b_{k-2},b_{k-1},b_k)$, if $k$ is odd,
and $f^{\prime}(b_{k-1},b_{k-2},b_{k-1},b_k)$ otherwise.
The probabilities of configurations of length $3$ can be
determined by requiring~$\sum_{b_0,b_1} q_{b_0 b_1 b_2 b_3 b_4}=q_{b_2 b_3 b_4}$.

The probability vector~\eqref{eq:tv} can be efficiently encoded within an MPA by
introducing a $3$-dimensional auxiliary space, with basis vectors
$\kket{0}$, $\kket{1}$, $\kket{2}$ keeping the relevant information about the
last three bits. The pairs of matrices $A_s(p_r,p_l)$, $A^{\prime}_s(p_r,p_l)$
and left/right boundary vectors $\bbra{L(p_r,p_l)}$, $\kket{R(p_r,p_l)}$ are
constructed, so that the probability vector $\vec{q}$ is expressed as
\begin{eqnarray}\label{eq:MPStimestate}
  \vec{q}=\bbra{L}\vec{A}_0\vec{A}^{\prime}_1\vec{A}_2\cdots\vec{A}^{(\prime)}_{k}\kket{R}.
\end{eqnarray}
The last matrix is either $\vec{A}$ or $\vec{A}^{\prime}$, depending on the parity of
the last index $k$. The matrices $A_s(p_r,p_l)$ contain the relevant transition
rates $f$, $g$,
\begin{eqnarray}
  A_0=
  \begin{bmatrix}
    1-p_r&0&0\\
    0&0&0\\
    1&0&0
  \end{bmatrix},\qquad
  A_1=
  \begin{bmatrix}
    0&p_r&0\\
    0&0&1\\
    0&0&0
  \end{bmatrix},
\end{eqnarray}
and the other pair of matrices is obtained by replacing $p_r$ with $p_l$,
\begin{eqnarray}
  \vec{A}^{\prime}(p_r,p_l)=\vec{A}(p_l,p_r).
\end{eqnarray}
The left and right vectors $\bbra{L}$, $\kket{R}$ are the left and
the right eigenvector of
the matrix~$\mathcal{T}=(A_0+A_1)(A_0^{\prime}+A_1^{\prime})$
that correspond to the leading eigenvalue (which is $1$), namely
\begin{eqnarray}
 \bbra{L}=\frac{1}{1+p_r+p_l}\begin{bmatrix}
    1&p_l&p_r
  \end{bmatrix},\qquad
  \kket{R}=\begin{bmatrix}
    1\\1\\1
  \end{bmatrix}.
\end{eqnarray}
Additionally, this choice of $\kket{R}$ works for both parities of $k$, since $\kket{R}$
is an eigenvector corresponding to the eigenvalue $1$ of both $A_0+A_1$ and $A_0^{\prime}+A_1^{\prime}$.
The normalization of the boundary vectors is determined by the normalization
condition of the probability vector~$\vec{q}$,
\begin{eqnarray}
    \smashoperator{\sum_{b_0,b_1,b_2,\ldots,b_{k}}} q_{b_0 b_1 b_2\ldots b_{k}}\!=\!
    \bbra{L}\overbrace{(A_0+A_1)(A_0^{\prime}+A_1^{\prime})(A_0+A_1)\cdots}^{k+1}
    \kket{R}\!=\!\bbrakket{L}{R}\!=\!1.
\end{eqnarray}

\section{Two-point correlation functions}\label{sec:2pointCorrs}
Since the matrices $\vec{A}$, $\vec{A}^{\prime}$ are finitely dimensional,
numerous physically interesting quantities can now be expressed explicitly.
The simplest example is the 2-point density-density correlation function at
different times but at the same position,
\begin{eqnarray}
  C(t) = \ave{\rho(0,0)\rho(0,t)}-\ave{\rho(0,0)}^2,
\end{eqnarray}
where $\rho(x,t)$ is the density at position $x$ and time $t$, and $\ave{\cdot}$
is the expectation value in the (horizontal) stationary state. The correlation
function is the~$x=0$ part of the spatio-temporal correlation function computed
in~\cite{TMPA2018}, but here the underlying (horizontal) equilibrium state is
richer and not limited to the maximum entropy (\emph{infinite temperature})
state. %considered in Ref.~\cite{TMPA2018}.

For simplicity, let us consider even times $t=2m$. Then, the correlation
function takes the following matrix-product form,
\begin{eqnarray}
  \eqalign{
    C(2m)&
    =\bbra{L}A_1 \overbrace{(A^{\prime}_0+A^{\prime}_1) (A_0+A_1)
    \cdots (A^{\prime}_0+A^{\prime}_1)}^{2m-1}A_1\kket{R}
  -\bbra{L}A_1\kket{R}^2\\
  &=\bbra{L}A_1(A^{\prime}_0+A^{\prime}_1)\mathcal{T}^{m-1}
  A_1\kket{R}-\frac{(p_l+p_r)^2}{(1+pl+pr)^2},
}
\end{eqnarray}
where $\mathcal{T}$ is a {\em temporal transfer matrix}, i.e.\ the product of
sums of both types of MPA
matrices,~$\mathcal{T}=(A_0+A_1)(A_0^{\prime}+A_1^{\prime})$, as introduced
earlier. The correlation function decays exponentially, with the exact form
given by
\begin{eqnarray}\fl
  C(2m)=\frac{
  (\Lambda_2^{m}-\Lambda_3^m)((p_l+p_r)^2\!-p_l p_r(1+p_l+p_r))
    +(\Lambda_2^{m-1}\!-\Lambda_3^{m-1})p_l p_r (p_l+p_r)
  }{(1+p_l+p_r)^2(\Lambda_3-\Lambda_2)},
\end{eqnarray}
where $\Lambda_{2,3}$ are the subleading eigenvalues of
the matrix~$\mathcal{T}$,\footnote{Note that this holds only
for $\Lambda_2\neq\Lambda_3$. If the subleading eigenvalues
are degenerate, the exact form of the correlation function is slightly
changed, but asymptotically the decay is still exponential.
}
\begin{eqnarray}
  \Lambda_{2,3}=-\frac{1}{2}\left(
    p_l+p_r-p_l p_r\pm\sqrt{(p_l+p_r-p_l p_r)^2-4 p_l p_r}
  \right).
\end{eqnarray}
In the maximum entropy state, i.e.\ $\xi=\omega=1$ or $p_l=p_r=\frac{1}{2}$,
the correlation function can be recast as
\begin{eqnarray}\fl
  C^{\infty}(2m) =
  2^{-3m-4}\left(
    \left(-1-\frac{\ii}{\sqrt{7}}\right)\left(-3-\ii\sqrt{7}\right)^m
    +\left(-1+\frac{\ii}{\sqrt{7}}\right)\left(-3+\ii\sqrt{7}\right)^m
  \right),
\end{eqnarray}
which agrees with the result of~\cite{TMPA2018}.

\section{Probabilities of time-configurations and the absence of decoupling of
timescales in RCA54}
\label{sec:timeConfProbs}

Even though the $2$-point correlation function is a generalization of the
previous result~\cite{TMPA2018}, the true advantage of the MPA is the ability
to easily express multi-point correlation functions of observables at one
specific location in space,
{as introduced in Eq.~\eqref{eq:expVal}.
  Explicitly, for~$t_1\le t_2\le t_3\ldots \le t_k$, the multi-time correlation
  function of one-site observables~$a_1, a_2,\ldots a_k$ can be expressed in
  terms of components of the time-state vector~$\vec{q}$ as
  \begin{eqnarray}\label{eq:expValExplicit}
    C_{a_1,a_2,\ldots,a_k}(t_1,t_2,\ldots,t_k)=
    \smashoperator{\sum_{s_0,s_1,s_2,\ldots,s_{t_k}}}
    q_{s_0 s_1 s_2\ldots s_{t_k}}
    a_1(s_{t_1}) a_2(s_{t_2})\cdots a_{t_k}(s_{t_k}),
  \end{eqnarray}
  where~$a_{j}(s_j)$ is the value of the observable~$a_j$, associated to
the one-site configuration~$s_j$.}

{ 
  The probabilities of time configurations can be understood as extreme cases
  of multi-time correlation functions, when only one of the terms in the
  sum~\eqref{eq:expValExplicit} has a nonzero prefactor. They can be
  alternatively recast in an equivalent (non-MPA) form as
}
%An extreme case are just the
%probabilities of time configurations themselves. These can be alternatively
%recast in an equivalent (non MPA) form as,
\begin{eqnarray}\label{eq:newTConfProb}
  q_{s_1,s_2,\cdots,s_k}=\delta_{n_1,0}
  \frac{(1-p_l)^{n_2}(1-p_r)^{n_3}p_r^{n_4}p_l^{n_5}}{1+p_l+p_r},
\end{eqnarray}
with $\{n_j,j=1,\ldots,5\}$ being a set of integer coefficients that
characterize the number of left and right movers in the configuration
$(s_1,s_2,\ldots,s_k)$,
\begin{eqnarray}\fl
  \eqalign{
  &\eqalign{
  n_1=\sum_{j=2}^{k-1} s_j \left(1-(s_{j-1}-s_{j+1})^2\right),\quad
  &n_2=\sum_{j=1}^{\lfloor k/2\rfloor} (1-s_{2j-1})(1-s_{2j}),\\
  n_3=\smashoperator{\sum_{j=1}^{\lfloor (k-1)/2\rfloor}} (1-s_{2j})(1-s_{2j+1}).
    &n_4=\smashoperator{\sum_{j=1}^{\lfloor k/2\rfloor}}
    s_{2j-1} s_{2j}
    + {\mathrm{mod}}(k+1,2) (1-s_{k-1}) s_k,
}\\
&n_5=s_1(1-s_2)+
    \smashoperator{\sum_{j=1}^{\lfloor (k-1)/2\rfloor}}
    s_{2j} s_{2j+1} + \mathrm{mod}(k,2) (1-s_{k-1}) s_k.
  }
\end{eqnarray}
The expression~\eqref{eq:newTConfProb} is equivalent to the appropriate
component of the time vector~\eqref{eq:MPStimestate}, which can be straightforwardly
checked.

In principle, the result above gives access to the full distribution of
timescales associated with the dynamics of a single site, when tracing out the
rest of the system. This information is particularly important in systems that
display complex dynamics and are therefore likely to exhibit non-trivial
distributions of local times. As a particular example we consider the
distributions of \emph{exchange} and \emph{persistence} times in the RCA54. The
definitions are those used in the literature on supercooled liquids;
see~\cite{Jung2004,Jung2005}. The exchange time is the time between two events,
for example a spin flip.%, after a previous event has occurred, i.e.\ it is the
%time between two events.
The persistence time is the time to the next event,
starting from an arbitrary time, i.e.\ not conditioned on the previous event
having occurred at the start of the time segment under consideration. 

If the statistics of events is {\em Poissonian}, persistence and exchange times
are equally (and exponentially) distributed. In general, however, when there
are non-trivial correlations between events, both types of times measure
different properties of the underlying process and their distributions differ.
For example, in stochastic kinetically constrained models (KCMs) of glasses,
like the Fredrickson-Andersen (FA) \cite{Fredrickson1984,Garrahan2018}, or the
East model \cite{Jackle1991,Garrahan2018}, the two timescales behave very
differently, with the average persistence time growing much faster with
the decreasing temperature than the average exchange time. This phenomenon is known as
{\em transport decoupling} \cite{Jung2004,Jung2005}. The reason is that the
dynamics of these stochastic KCMs is intermittent, and when projected to a
single site --- in analogy to what is done here for the RCA54 cellular
automaton --- it leads to the \emph{bunching of flip events} on the site. In this
case, the exchange
time is dominated by the bunched events, while the persistence time is
dominated by the long stretches between bunches. 
It is interesting to consider the possibility of timescale decoupling in the
RCA54, as the dynamical rule~\eqref{eq:timeProp}
imposes a constraint that is similar to that of the FA model (flips can only
occur if at least one nearest neighbour is excited). Furthermore, the RCA54
displays dynamical large-deviation transitions \cite{bucaetalLargeDev} similar to those of
the FA model \cite{Garrahan2007}, which indicates the presence of non-trivial
fluctuations in the trajectories of the dynamics. 

For the RCA54 we define the exchange time  as the time between two
consecutive observed particles (excitations) in a time configuration, and the
persistence time as the time before the next particle, if we start observing the
state at some point. More precisely, we can define
the distribution functions of exchange and
persistence times and compute them from our explicit form of the time vector in
the RCA54. The probability of observing an~exchange time equal to $t$,
$p_{\text{E}}(t)$, can be simply expressed in terms of probabilities of time
configurations of the form $100\ldots01$,
\begin{eqnarray}
    p_{\text{E}}(t)\propto \frac{1}{2}\left(
    q_{1\scriptstyle{\underbrace{\scriptstyle{00\ldots0}}_{t}}1}+
    q_{01\scriptstyle{\underbrace{\scriptstyle{00\ldots0}}_{t}}1}+
    q_{11\scriptstyle{\underbrace{\scriptstyle{00\ldots0}}_{t}}1}
  \right),\qquad \sum_{t=1}^{\infty} p_{\text{E}}(t)=1,
\end{eqnarray}
where we average over the even/odd starting positions of the time
configuration. Note that the second term can be dropped, since the subsequence $010$
is forbidden.  Similarly, the probability of persistence time being equal to
$t$, $p_{\text{P}}(t)$, is, up to normalization, equal to
\begin{eqnarray}
  p_{\text{P}}(t)\propto \frac{1}{2}\left(
    q_{\scriptstyle{\underbrace{\scriptstyle{00\ldots0}}_{t}}1}+
    q_{0\scriptstyle{\underbrace{\scriptstyle{00\ldots0}}_{t}}1}+
    q_{1\scriptstyle{\underbrace{\scriptstyle{00\ldots0}}_{t}}1}
  \right),\qquad \sum_{t=1}^{\infty} p_{\text{P}}(t)=1.
\end{eqnarray}
The two distributions are related by 
\begin{eqnarray}
  p_{\text{P}}(t)-p_{\text{P}}(t+1) = \frac{p_{\text{E}}(t)}{\ave{t_{\rm E}}},
\end{eqnarray}
which is the discrete-time version of the usual relation between persistence
and exchange distributions in continuous-time dynamics. Here,
$\ave{t_{\rm E}}$, $\ave{t_{\rm P}}$ are the average exchange and persistence times;
$\ave{t_{\rm E/P}}=\sum_{t=1}^{\infty} t\,p_{_{\rm E/P}}(t)$.

Using the~simplified expressions for probabilities~\eqref{eq:newTConfProb},
it is easy to see that the exchange and persistence time distributions
reduce to the following form:
\begin{eqnarray}
  \eqalign{
    p_{\text{E}}(t)= \begin{cases}
    \frac{p_l^2(1-p_r)+p_r^2(1-p_l)}{p_l+p_r}
    \left((1-p_l)(1-p_r)\right)^{t/2-1}
    ,\quad & t\equiv0\pmod{2},\\
    \frac{2 p_l p_r}{p_l+p_r}
    \left((1-p_l)(1-p_r)\right)^{(t-1)/2}
    , & t\equiv1\pmod{2},
  \end{cases}\\
  p_{\text{P}}(t)=\begin{cases}
    \frac{p_l(1-p_r)+p_r(1-p_l)}{2}
    \left((1-p_l)(1-p_r)\right)^{t/2-1}
    ,\quad & t\equiv0\pmod{2},\\
    \frac{p_l+p_r}{2}
    \left((1-p_l)(1-p_r)\right)^{(t-1)/2}
    , & t\equiv1\pmod{2}.
  \end{cases}
}
\end{eqnarray}
If the parameters $p_l$, $p_r$ coincide, the two distributions are the same.
Otherwise they differ, but the deviations from the Poissonian
statistics are small and can be understood as a~consequence of short-range
correlations of the time state. Specifically, computing the average exchange
and persistence times,
\begin{eqnarray}
  \ave{t_{\rm E}} = \frac{2}{p_l+p_r},\qquad
  \ave{t_{\rm P}} = \frac{4-p_l-p_r}{2(p_l+p_r-p_l p_r)},
\end{eqnarray}
we observe that their ratio is bounded from top and bellow as
\begin{eqnarray}
  1\le\frac{\ave{t_{\rm E}}}{\ave{t_{\rm P}}} \le \frac{4}{3},
\end{eqnarray}
which indicates that, overall, the two timescales scale similarly with the
parameters that control the dynamics, in contrast to what occurs in KCMs like
the FA model \cite{Jung2004,Jung2005}. This also means that, in contrast to
stochastic KCMs, to quantify non-trivial correlations in the dynamics of the
RCA54 (cf.~\cite{bucaetalLargeDev}), it is necessary to consider multi-point
correlators both in time and space (not just in time as above). 

\section{Conclusion}\label{sec:conclusion}

We have constructed an explicit matrix-product form of the time state~$\vec{q}$
in the case of rule 54 reversible cellular automaton. The construction was
possible due to the simple structure of the equilibrium state~$\vec{p}$, which
contains left and right movers that are statistically independent, as long as
they follow an exclusion rule. The MPA form of the vector $\vec{q}$ allows us
to efficiently express correlation functions and expectation values of time
configurations, i.e.\ multi-time correlation functions. For instance, we have
computed $2$-point correlation function and shown that, in the limit
$\xi=\omega=1$ (i.e.\ in the maximum entropy state), it agrees with the result of
Ref~\cite{TMPA2018}. Furthermore, it is possible to express the probabilities
of time configurations in terms of numbers of left and right movers, which
enables us to exactly express the distributions of exchange and persistence
times and in this way show the absence of timescale decoupling in the model.

The work presented in the paper opens some new research questions. The first
one is the generalization of the result to other time-translation invariant
(equilibrium) states.  The model exhibits an infinite number of conserved
quantities, therefore it is possible to construct states~$\vec{p}$ that are
characterized by more than $2$ parameters. The question is whether in this
case it is still possible to obtain an efficient MPA representation of the time
state and what is the dimensionality of the necessary auxiliary space.

Even though the construction of the time state is straightforward, one would like
to obtain some algebraic interpretation, for example by finding relations similar to
the cubic algebra in~\cite{prosenBucaCA54,bucaetalLargeDev}. They could provide
simpler means of extracting physically interesting quantities, as well as
contribute additional insight into the integrability structure of the model.

Most prominently, one would like to generalize the result or apply the concept 
to other lattice dynamical systems. 
This is intimately connected to the previous question of understanding the solution
from an algebraic point of view, which is more robust and does not rely on
particularities of the model and the underlying stationary state.  Preliminary empirical
results suggest that generically, the computational complexity of the time
state grows exponentially with time,  however one could hope to devise an
approximate numerical scheme to efficiently simulate the dynamics (time state)
at $x=0$ using the MPA as a variational ansatz~\cite{schollwock2011density}.

\ack
The authors thank Lenart Zadnik for the useful comments on the manuscript.
This work has been supported by the European Research Council under the
Advanced Grant No.\ 694544 - OMNES, by the Slovenian Research Agency(ARRS)
under the Programme P1-0402, and by the Leverhulme Trust Grant No.\ RPG-2018-181. 

\section*{References}
\bibliographystyle{iopart-num}
\bibliography{bobenko}

\providecommand{\newblock}{}
\begin{thebibliography}{10}
\expandafter\ifx\csname url\endcsname\relax
  \def\url#1{{\tt #1}}\fi
\expandafter\ifx\csname urlprefix\endcsname\relax\def\urlprefix{ }\fi
\providecommand{\href}[2]{#1}  % in case hyperref not loaded
% \eprint[archive=arXiv]{identifier}
\providecommand{\eprint}[2][arXiv]{#1:\linebreak[0]#2}
% Bibliography created with iopart-num v2.1+

\bibitem{Chandler2010}
Chandler D and Garrahan J~P 2010 {\em Annu. Rev. Phys. Chem.\/} {\bf 61}
  191--217
  \urlprefix\url{http://dx.doi.org/10.1146/annurev.physchem.040808.090405}

\bibitem{Binder2011}
Binder K and Kob W 2011 {\em Glassy materials and disordered solids: An
  introduction to their statistical mechanics\/} (Singapore: World Scientific)

\bibitem{Biroli2013}
Biroli G and Garrahan J~P 2013 {\em J. Chem. Phys.\/} {\bf 138} 12A301
  \urlprefix\url{http://link.aip.org/link/?JCP/138/12A301/1}

\bibitem{spohn2014nonlinear}
Spohn H 2014 {\em J. Stat. Phys.\/} {\bf 154} 1191--1227
  \urlprefix\url{https://link.springer.com/article/10.1007/s10955-014-0933-y}

\bibitem{bobenko1993two}
Bobenko A, Bordemann M, Gunn C and Pinkall U 1993 {\em Commun. Math. Phys.\/}
  {\bf 158} 127--134 \urlprefix\url{https://doi.org/10.1007/BF02097234}

\bibitem{Takesue}
Takesue S 1987 {\em Phys. Rev. Lett.\/} {\bf 59}(22) 2499--2502
  \urlprefix\url{https://link.aps.org/doi/10.1103/PhysRevLett.59.2499}

\bibitem{gopalakrishnan1}
Gopalakrishnan S 2018 {\em Phys. Rev. B\/} {\bf 98}(6) 060302
  \urlprefix\url{https://link.aps.org/doi/10.1103/PhysRevB.98.060302}

\bibitem{gopalakrishnan2}
Gopalakrishnan S, Huse D~A, Khemani V and Vasseur R 2018 {\em Phys. Rev. B\/}
  {\bf 98}(22) 220303
  \urlprefix\url{https://link.aps.org/doi/10.1103/PhysRevB.98.220303}

\bibitem{gopalakrishnan3}
Friedman A~J, Gopalakrishnan S and Vasseur R 2019 {\em Phys. Rev. Lett.\/} {\bf
  123}(17) 170603
  \urlprefix\url{https://link.aps.org/doi/10.1103/PhysRevLett.123.170603}

\bibitem{prosenMejiaMonasterioCA54}
Prosen T and Mej{\'\i}a-Monasterio C 2016 {\em J. Phys. A: Math. Theor.\/} {\bf
  49} 185003 \urlprefix\url{http://stacks.iop.org/1751-8121/49/i=18/a=185003}

\bibitem{inoueTakesueCA54}
Inoue A and Takesue S 2018 {\em J. Phys. A: Math. Theor.\/} {\bf 51} 425001
  \urlprefix\url{https://doi.org/10.1088%2F1751-8121%2Faadc29}

\bibitem{prosenBucaCA54}
Prosen T and Bu\v{c}a B 2017 {\em J. Phys. A: Math. Theor.\/} {\bf 50} 395002
  \urlprefix\url{http://stacks.iop.org/1751-8121/50/i=39/a=395002}

\bibitem{bucaetalLargeDev}
{Bu{\v{c}}a} B, {Garrahan} J~P, {Prosen} T and {Vanicat} M 2019 {\em Phys. Rev.
  E\/} {\bf 100}(2) 020103
  \urlprefix\url{https://link.aps.org/doi/10.1103/PhysRevE.100.020103}

\bibitem{TMPA2018}
Klobas K, Medenjak M, Prosen T and Vanicat M 2019 {\em Commun. Math. Phys.\/}
  {\bf 371} 651--688 \urlprefix\url{https://doi.org/10.1007/s00220-019-03494-5}

\bibitem{alba2019RCA54}
Alba V, Dubail J and Medenjak M 2019 {\em Phys. Rev. Lett.\/} {\bf 122}(25)
  250603
  \urlprefix\url{https://link.aps.org/doi/10.1103/PhysRevLett.122.250603}

\bibitem{Jung2004}
Jung Y, Garrahan J and Chandler D 2004 {\em Phys. Rev. E\/} {\bf 69}
  \urlprefix\url{http://dx.doi.org/061205}

\bibitem{Jung2005}
Jung Y, Garrahan J~P and Chandler D 2005 {\em J. Chem. Phys.\/} {\bf 123}
  084509 \urlprefix\url{https://aip.scitation.org/doi/abs/10.1063/1.2001629}

\bibitem{Fredrickson1984}
Fredrickson G~H and Andersen H~C 1984 {\em Phys. Rev. Lett.\/} {\bf 53}
  1244--1247
  \urlprefix\url{https://link.aps.org/doi/10.1103/PhysRevLett.53.1244}

\bibitem{Garrahan2018}
Garrahan J~P 2018 {\em Physica A\/} {\bf 504} 130--154
  \urlprefix\url{https://doi.org/10.1016/j.physa.2017.12.149}

\bibitem{Jackle1991}
J{\"a}ckle J and Eisinger S 1991 {\em Z. fur Phys. B\/} {\bf 84} 115--124 ISSN
  0722-3277 \urlprefix\url{http://dx.doi.org/10.1007/BF01453764}

\bibitem{Garrahan2007}
Garrahan J~P, Jack R~L, Lecomte V, Pitard E, van Duijvendijk K and van Wijland
  F 2007 {\em Phys. Rev. Lett.\/} {\bf 98}(19) 195702
  \urlprefix\url{http://dx.doi.org/10.1103/PhysRevLett.98.195702}

\bibitem{schollwock2011density}
Schollw{\"o}ck U 2011 {\em Ann. Phys.\/} {\bf 326} 96--192
  \urlprefix\url{https://doi.org/10.1016/j.aop.2010.09.012}

\end{thebibliography}
\appendix
\section{Boundary driving with the steady state that corresponds to the
asymptotic probability distribution on a finite chain}\label{app:boundaryProps}
We would like to connect the MPA form of the vectors $\vec{p}$,
$\vec{p}^{\prime}$, introduced in~\eqref{eq:asymptoticDistrs}, with an
equivalent MPA solution of a boundary driven setup. This amounts to
constructing boundary two-site propagators $P^{L/R}$, so that
$U_{\text{e/o}}$, defined as
\begin{eqnarray}\fl
  U_{\text{e}}=P_{1 2 3} P_{3 4 5} \cdots P_{2m-3\,2m-2\,2m-1} P^R_{2m-1\,2m},\qquad
  U_{\text{o}}=P^L_{1 2} P_{2 3 4} P_{4 5 6} \cdots P_{2m-2\,2m-1\,2m},
\end{eqnarray}
map $\vec{p}$ to $\vec{p}^{\prime}$, and vice versa.

Taking into account the bulk relations~\eqref{eq:cubicRelation}
and~\eqref{eq:dualCubicRelation}, it suffices that the following boundary
equations are satisfied,
\begin{eqnarray}\fl \label{eq:boundaryRelations}
  \eqalign{
    \bra{l} \vec{W}_1 S = \alpha \bra{l^{\prime}}\vec{W}_1^{\prime},
    \qquad
    &P^L_{12}\left(\bra{l^{\prime}}\vec{W}^{\prime}_1 \vec{W}_2 S\right)
    =\gamma\bra{l}\vec{W}_1\vec{W}^{\prime}_2,\\
    \vec{W}^{\prime}_{2m} S \ket{r^{\prime}} = \delta \vec{W}^{\prime}_{2m}\ket{r},
    \qquad
    &P^R_{2m-1\,2m}\left(\vec{W}^{\prime}_{2m-1} S \vec{W}^{\prime}_{2m}\ket{r}\right)
    =\beta \vec{W}^{\prime}_{2m-1}\vec{W}_{2m}\ket{r^{\prime}},
  }
\end{eqnarray}
where the parameters $\alpha$, $\beta$, $\gamma$ and $\delta$ have to fulfill
the condition
\begin{eqnarray}
  \alpha \beta =\frac{\braket{l}{r}}{\braket{l^{\prime}}{r^{\prime}}} = \frac{1}{\gamma  \delta}.
\end{eqnarray}
Solving these equations, we obtain the following boundary propagators,
\begin{eqnarray}\fl
  P^R=\begin{bmatrix}
    1-p_l&1-p_l&0&0\\
    p_l&p_l&0&0\\
    0&0&0&1\\
    0&0&1&0
  \end{bmatrix},\qquad
  P^L=\begin{bmatrix}
    1-p_r&0&1-p_r&0\\
    0&0&0&1\\
    p_r&0&p_r&0\\
    0&1&0&0
  \end{bmatrix},
\end{eqnarray}
with parameters $p_r$, $p_l$ corresponding to the conditional probabilities of observing
right/left movers as expressed in~\eqref{eq:probRight}, \eqref{eq:probLeft}, and the following
values of $\alpha$, $\beta$, $\gamma$, $\delta$,
\begin{eqnarray}
  \eqalign{
    \alpha=\frac{\lambda+\xi-\xi\omega}{\lambda+\omega-\xi\omega},\qquad &\beta=\frac{\omega}{\xi},\\
    \gamma=1,
    \qquad&\delta=\frac{\xi(\lambda+\omega-\xi\omega)}{\omega(\lambda+\xi-\xi \omega)}.
  }
\end{eqnarray}
Note that the choice of $P^{R,L}$ is not unique.

The boundary relations~\eqref{eq:boundaryRelations} and the
solution~\eqref{eq:asymptoticDistrs} are simpler than the ones considered
in~\cite{prosenBucaCA54}, since the boundary vectors do not depend on the
configuration of the boundary sites. These states therefore represent a
specific subset of solutions found in the previous
works~\cite{prosenMejiaMonasterioCA54,inoueTakesueCA54,prosenBucaCA54}.

\section{Conditional probabilities of observing subconfigurations}\label{app:condProbs}
To prove the relations~\eqref{eq:cond4site1} and~\eqref{eq:cond4site2}, we
observe that the vectors
$\bra{l} W_{s_1} W^{\prime}_{s_2} W_{s_3} W^{\prime}_{s_4}$ and $\bra{l} W_{s_3} W^{\prime}_{s_4}$
(similarly $W_{s_1} W^{\prime}_{s_2} W_{s_3} W^{\prime}_{s_4} \ket{r}$ and $W_{s_1} W^{\prime}_{s_2}\ket{r}$) 
are linearly dependent. Explicitly, for every $4$-site configuration $(s_1,s_2,s_3,s_4)$
there exist scalar factors $c^{(l)}_{s_1 s_2 s_3 s_4}$ and
$c^{(r)}_{s_1 s_2 s_3 s_4}$, so that the following holds,
\begin{eqnarray}\label{eq:linearDependenceL}
    \bra{l} W_{s_1} W^{\prime}_{s_2} W_{s_3} W^{\prime}_{s_4}
    = c^{(l)}_{s_1 s_2 s_3 s_4} \bra{l} W_{s_3} W^{\prime}_{s_4},\\
    W_{s_1}W^{\prime}_{s_2}W_{s_3}W^{\prime}_{s_4}\ket{r} = c^{(r)}_{s_1 s_2 s_3 s_4}
    W_{s_1} W^{\prime}_{s_2}\ket{r},\label{eq:linearDependenceR}
\end{eqnarray}
which can be straightforwardly verified by checking all $16$ configurations.
From here relations~\eqref{eq:cond4site1} and~\eqref{eq:cond4site2} follow directly, since
for every even configuration length $2 k$ the left-hand side of the first part of equation~\eqref{eq:cond4site1}
can be rewritten as
\begin{eqnarray}
  \eqalign{
  \frac{\bra{l} W_{s_1}W^{\prime}_{s_2}\cdots W_{s_{2k-3}}W^{\prime}_{s_{2k-2}}W_{s_{2k-1}}W^{\prime}_{s_{2k}}\ket{r}}{\bra{l} W_{s_1}W^{\prime}_{s_2}\cdots W_{s_{2k-3}}W^{\prime}_{s_{2k-2}}\ket{r}}\\
  =\frac{c^{(l)}_{s_1 s_2 s_3 s_4}\cdots
        %c^{(l)}_{s_3 s_4 s_5 s_6} \cdots
        c^{(l)}_{s_{2k-5} s_{2k-4} s_{2k-3} s_{2k-2}} 
    \bra{l}W_{s_{2k-3}}W^{\prime}_{s_{2k-2}}W_{s_{2k-1}} W^{\prime}_{s_{2k}}\ket{r}}
  {c^{(l)}_{s_1 s_2 s_3 s_4} \cdots
        %c^{(l)}_{s_3 s_4 s_5 s_6} \cdots
        c^{(l)}_{s_{2k-5} s_{2k-4} s_{2k-3} s_{2k-2}} 
  \bra{l}W_{s_{2k-3}}W^{\prime}_{s_{2k-2}}\ket{r}}\\
=\frac{\bra{l}W_{s_{2k-3}}W^{\prime}_{s_{2k-2}}W_{s_{2k-1}} W^{\prime}_{s_{2k}}\ket{r}}
  {\bra{l}W_{s_{2k-3}}W^{\prime}_{s_{2k-2}}\ket{r}},
}
\end{eqnarray}
where the first equality follows from the repeated application of
the relation~\eqref{eq:linearDependenceL}.
From Eq.~\eqref{eq:linearDependenceR} we analogously obtain
\begin{eqnarray}\fl
  \frac{\bra{l} W_{s_1}W^{\prime}_{s_2}W_{s_3}W^{\prime}_{s_4}\cdots W_{s_{2k-1}}W^{\prime}_{s_{2k}}\ket{r}}{\bra{l} W_{s_3}W^{\prime}_{s_4}\cdots W_{s_{2k-1}}W^{\prime}_{s_{2k}}\ket{r}}=
  \frac{\bra{l}W_{s_{1}}W^{\prime}_{s_{2}}W_{s_{3}} W^{\prime}_{s_{4}}\ket{r}}
  {\bra{l}W_{s_{3}}W^{\prime}_{s_{4}}\ket{r}}.
\end{eqnarray}
Note that the pair of relations dealing with probabilities
$p^{\prime}_{\underline{s}}$ follows trivially after the exchange
$\xi\leftrightarrow \omega$.
\end{document}